\documentstyle[aps,twocolumn,epsfig,tighten,floats]{revtex}
\begin{document}
\newcommand{\tr}{\mbox{tr}\,}
\newcommand{\eq}[1]{{Eq.~(\ref{#1})}}
\newcommand{\fig}[1]{{Fig.~\ref{#1}}}
\newcommand{\tab}[1]{{Table~\ref{#1}}}
\draft
\title
{\hfill\begin{minipage}{0pt}\scriptsize \begin{tabbing}
\hspace*{\fill} GUTPA/00/11/01\\
\hspace*{\fill} WUB 01-01\end{tabbing}
\end{minipage}\\[8pt]
Quark mass effects on the topological
susceptibility in QCD}
\author{Gunnar S.\ Bali$^a$, Norbert Eicker$^b$,
Thomas Lippert$^b$,
Hartmut Neff$^c$,
Boris Orth$^b$,\\
Klaus Schilling$^{bc}$,
Thorsten Struckmann$^c$
and Jochen Viehoff$^c$}
\address{$^a$Department of Physics \& Astronomy, The University of Glasgow, Glasgow G12 8QQ, Scotland}
\address{$^b$Fachbereich Physik, Bergische Universit\"at
Wuppertal, D-42097 Wuppertal, Germany}
\address{$^c$von Neumann Institute for Computing,
c/o Research Center J\"ulich, D-52425 J\"ulich and DESY, D-22603 Hamburg, Germany}
\author{(SESAM and T$\chi$L Collaborations)}
\date{\today}
\maketitle

\begin{abstract}
  We investigate topology in lattice simulations of QCD with two
  flavours of dynamical Wilson fermions.  At various sea quark masses
  we find reasonable agreement between results for the topological
  charge from fermionic and gluonic definitions, the latter with
  cooling.  We study correlations of the topological charge with light
  hadronic observables and find effects in the flavour singlet
  pseudoscalar channel. The quark mass dependence of the
  topological susceptibility, $\chi$, is consistent with the
  leading order small $m_{\pi}$ expectation,
  $\chi=f_{\pi}^2m_{\pi}^2/4$.
\end{abstract}

\narrowtext
\section{Introduction}
One of the most intriguing features of QCD is its topological vacuum
structure which results in phenomenological consequences with an
important bearing on particle physics such as the breaking of the
axial $U_A(1)$ and chiral symmetries.

In the context of {\it pure gauge} theories, lattice methods have by
now matured to provide a valuable tool kit for accessing the continuum
topological susceptibility, $\chi$.  Various gluonic definitions of
the topological charge lead to consistent estimates of
$\chi$~\cite{Alles:1998nu}, and remnants of the Atiyah-Singer index
theorem are witnessed to hold on the lattice: the topological charge,
$Q$, as obtained after cooling from the gluonic degrees of freedom is
consistent with the fermionic index, as determined from counting the
zero-level crossings of the eigenvalues of the Wilson-Dirac operator,
$\gamma_5(D\!\!\!\!/+m_0)$, under variations of the bare quark mass,
$m_0\approx 0$~\cite{Itoh:1987iy,Narayanan:1997sa}.
Moreover, one might perceive the very value of $\chi$ as delivered by quenched
lattice
simulations~\cite{Teper:2000wp,Alles:1997nm,Hasenfratz:1998qk,GarciaPerez:1999ru}, $\chi_q^{1/4}=213(14)$~MeV,
to confirm the lattice approach to
topological properties: it turns out to be in fairly good agreement with the
large $N_c$ anticipation made by Witten and Veneziano~\cite{Witten:1979vv},
$\chi_q\approx f_{\pi}^2(m^2_{\eta'}+m_\eta^2-2m_K^2)/6\approx
(180\,\mbox{MeV})^4$.
In conclusion, the lattice machinery appears to work for the study of
topological aspects in gluodynamics.

Unfortunately, however, the situation regarding the
QCD vacuum proper is by far less settled:
for sheer cost reasons so far only the regime of
intermediate sea quark masses,
not much lighter than the strange quark, has been explored.
In fact it is still under
debate whether the data from full QCD simulations support the
expected {\it vanishing susceptibility} at chiral sea quark
masses, $m\rightarrow 0$:
%%%%%%%%%%%%
%in the small quark mass regime, the number
%of instantons $\nu_-$ will approach the number of anti-instantons
%$\nu_+$, resulting in a reduction of the modulus of the topological
%charge $|Q|=|\nu_--\nu_+|$ and, thus, a vanishing topological
%susceptibility
%%%%%%%%%%%%
\begin{equation}
\label{eq:chiral}
\chi=\frac{m}{n_f}\Sigma+{\mathcal O}(m^2),\quad
\Sigma=-\lim_{m\rightarrow 0}\lim_{V\rightarrow\infty}
\langle\bar{\psi}\psi\rangle.
\end{equation}
Attempts to verify this prediction via lattice simulations have a long
history. While the first lattice studies in this direction with Kogut-Susskind
(KS) sea quarks~\cite{Smit:1987jh,Gausterer:1989tg,Bitar:1991wr,mmp} could not
reach conclusive results, later investigations did yield some qualitative
evidence in favour of the expected decrease of
$\chi(m)$~\cite{Kuramashi:1993mv}.

Quite recently, this problem has been revisited from three different
sides --- but the debate is still open: the
CP-PACS~\cite{AliKhan:2000zi} and UKQCD~\cite{Hart:2000hy}
collaborations employed improved Wilson fermionic and different
gluonic actions while the Pisa group~\cite{Alles:2000cg} operated with
two and four flavours of KS fermions. A further data point
for $n_f=2$ KS flavours has been obtained by
Hasenfratz~\cite{Hasenfratz:2000ng}
(for a recent review
see e.g.\ Ref.~\cite{GarciaPerez}). While both the Pisa group
(working at an inverse lattice spacing $a^{-1}\approx 2$~GeV and bare
quark masses $am_0\geq 0.01$) and the CP-PACS collaboration (at
$a^{-1}\approx 1.3$~GeV, $m_{\pi}/m_{\rho}>0.59$) {\it saw no}
evidence whatsoever in favour of the expected chiral behaviour, the
UKQCD collaboration {\it did verify} (at $a^{-1}\approx 2$~GeV and
$m_{\pi}/m_{\rho}>0.57$) a decrease of $\chi (m)$, consistent with
theoretical expectations.

In this paper we shall present an analysis focused on this issue,
based on the final statistics of our SESAM and T$\chi$L samples of QCD
vacuum configurations.  The article is organised as follows. In
Sect.~\ref{sec:lat} we describe the details of our simulation and
methodology, including a comparison between fermionic and gluonic
definitions of the topological charge. Preliminary results on this
comparison, based on smaller statistical samples, have been reported
by us previously~\cite{Alles:1998jq}.  In Sect.~\ref{sec:hadron} we
investigate correlations between the topological charge and the hadron
spectrum.  Finally, in Sect.~\ref{sec:results}, we present the lattice
data on the topological susceptibility.

\section{Determination of the topological charge}
\label{sec:lat}
\subsection{Measurements}
We analyse ensembles of gauge configurations that have been generated
by means of the hybrid Monte Carlo (HMC) algorithm using the Wilson
fermionic and gluonic actions with $n_f=2$ mass degenerate quark
flavours at the inverse lattice coupling, $\beta=5.6$, corresponding
to an inverse lattice spacing $a^{-1}=(2.65^{+5}_{-8}\pm 0.14)$~GeV at
physical sea quark masses.  This was done on $L_{\sigma}^3\times
L_{\tau} =16^3\times 32$ as well as on $24^3\times 40$ lattices at
five different values of the sea quark mass parameter. The
corresponding chiralities can be quantified in terms of
$m_{\pi}/m_{\rho}$-ratios, ranging between $0.834(3)$ and
$0.574(13)$~\cite{inprep}.  The relevant simulation settings are
displayed in Table~\ref{tab:simul}~\cite{Bali:2000vr}.  At each
$\kappa$-value 4,000--5,000 thermalized HMC trajectories have been
generated.  In addition to the dynamical quark simulations, quenched
reference measurements on $16^4$ lattices at $\beta=6.0$ were
performed.  The configurations that enter the analysis are separated
by 25 (24 at $\kappa=0.1565$ and 10 in the quenched simulation) HMC
trajectories.

\begin{table}
\caption{Simulation parameters. Unless an error is stated
the autocorrelation times $\tau_{\mbox{\tiny int}}$ are lower estimates.
The number of measurements is related to the number of
trajectories by $n_{\mbox{\tiny meas}}=
\nu_{\mbox{\tiny meas}} n_{\mbox{\tiny traj}}$,
with $\nu_{\mbox{\tiny meas}}=1/10$ in the quenched
case ($\beta=6.0$), $\nu_{\mbox{\tiny meas}}=1/24$
at $\kappa=0.1565$, $\nu_{\mbox{\tiny meas}}=1/2$
in the second series of measurements at $\kappa=0.1575$, $L_{\sigma}=16$
and $\nu_{\mbox{\tiny meas}}=1/25$ otherwise.
The physical units have been obtained by setting $r_0^{-1}=394$~MeV
and are subject to scale uncertainties of about 5~\%. All unquenched
simulations have been performed at $\beta=5.6$.}
\label{tab:simul}
\begin{tabular}{cccccccc}
$\kappa$&
$L_{\sigma}$&
$n_{\mbox{\tiny meas}}$&
$r_0a^{-1}$&
$a_{\mbox{\tiny eff}}^{-1}/$GeV&
$m_{\pi}aL_{\sigma}$&
$m_{\pi}/m_{\rho}$&
$\tau_{\mbox{\tiny int}}$\\
\hline
0.1560&$16$&206&5.11(3)&2.01(1)&7.14(4)&0.834(3)&$30$\\
0.1565&$16$&209&5.28(5)&2.08(2)&6.39(6)&0.813(9)&$55$\\
0.1570&$16$&200&5.48(7)&2.16(3)&5.51(4)&0.763(6)&$40$\\
0.1575&$16$&215&5.96(8)&2.35(3)&4.50(5)&0.692(10)&$55$\\
 '' & '' &2200& '' & '' & '' & '' &54(4)\\
0.1575&$24$&150&5.89(3)&2.32(1)&6.65(6)&0.704(5)&$140$\\
0.1580&$24$&140&6.23(6)&2.45(2)&4.77(7)&0.574(13)&$130$\\\hline
$\kappa_{ph}$&---&---&$6.73^{+13}_{-19}$&$2.65^{+5}_{-8}$&---&0.179(1)&---\\\hline
$\beta=6.0$&$16$&320&5.33(3)&2.10(1)&---&---&$35$
\end{tabular}
\end{table}

Lower limits on the integrated autocorrelation times, $\tau_{\mbox{\scriptsize
    int}}$, are estimated by binning the data for the topological charge
$Q_i$, $i=1,\cdots,n_{\mbox{\scriptsize meas}}$, into $n_{\mbox{\scriptsize
    bin}}$ blocks that contain $m$ successive measurements each. On each such
block of length $m$ an average $\overline{Q}^m_j,
j=1,\ldots,n_{\mbox{\scriptsize bin}}$, is calculated:
$\overline{Q}^m_j=(1/m)\sum_{i=1}^{m}Q_{m(j-1)+i}$. We determine fluctuations
between these bins,
\begin{equation}
\Delta Q_m^2=\frac{1}{n_{\mbox{\scriptsize bin}}(n_{\mbox{\scriptsize bin}}-1)}
\sum_{j=1}^{n_{\mbox{\tiny bin}}}\left(
\overline{Q}^m_j-\langle Q\rangle\right)^2,
\end{equation}
where the average $\langle Q\rangle\approx 0$ is calculated on
the first
$n_{\mbox{\scriptsize bin}}\times m$ configurations.
We then estimate the autocorrelation times,
\begin{equation}
\tau_{\mbox{\scriptsize int}}\approx \frac{1}{2\,\nu_{\mbox{\scriptsize meas}}}
\,\,\max_{m^2\leq n_{\mbox{\tiny meas}}/2}\,
\left(\frac{\Delta Q_m^2}{\Delta Q_1^2}\right),
\end{equation}
that are included in the table.  $\nu_{\mbox{\scriptsize meas}}$
denotes the measurement frequency.  On the $16^3\times 32$ volume at
$\kappa=0.1575$, in addition to the $\nu_{\mbox{\scriptsize
    meas}}=1/25$ time series, we also determined the topological
charge with increased frequency $\nu_{\mbox{\scriptsize meas}}=1/2$,
with a reduced number of 10 (as opposed to 60) cooling sweeps.  The
larger frequency enabled us to compute the autocorrelation time from
the autocorrelation function itself, with full control over
statistical errors~\cite{Alles:1998jq,inprep2}.  The resulting value,
$\tau_{\mbox{\scriptsize int}}=54(4)$, is consistent with our estimate
obtained in the way described above, $\tau_{\mbox{\scriptsize
    int}}\approx 55$.  We take this as an indication that our
estimates are reasonable.  Interestingly, the autocorrelation times
grow both with decreasing quark mass and increasing volume as
already observed in Ref.~\cite{Alles:1998jq}.

We employ a gluonic as well as a fermionic definition of the topological
charge.  In the gluonic case we define a topological charge density,
\begin{equation}
q(x)=\frac{1}{16\pi^2}\tr F_{\mu\nu}(x)\tilde{F}_{\mu\nu}(x)
=\frac{1}{16\pi^2}F(x)*F(x),
\label{eq:gluonic_def_d}
\end{equation}
where we use the symmetric (clover leaf) definition~\cite{DiVecchia:1981qi},
\begin{equation}
F_{\mu\nu}=\frac{i}{4a^2}
\left(U_{\mu\nu}+U_{\nu,-\mu}+U_{-\mu,-\nu}+U_{-\nu,\mu}-4\right),
\end{equation}
and $\tilde{F}_{\mu\nu}=\frac{1}{2}\epsilon_{\mu\nu\rho\sigma}F_{\rho\sigma}$,
which is correct up to order $a^2$ lattice artefacts.
The topological charge,
\begin{equation}
Q=\sum_xq(x)=\frac{1}{16\pi^2}(F,*F),
\label{eq:gluonic_def}
\end{equation}
should then approach integer values as $a\rightarrow 0$ in the infinite volume
limit, on sufficiently smooth gauge configurations.

\subsection{Renormalization}
\begin{figure}[!htb]
\centerline{\epsfxsize=8cm\epsfbox{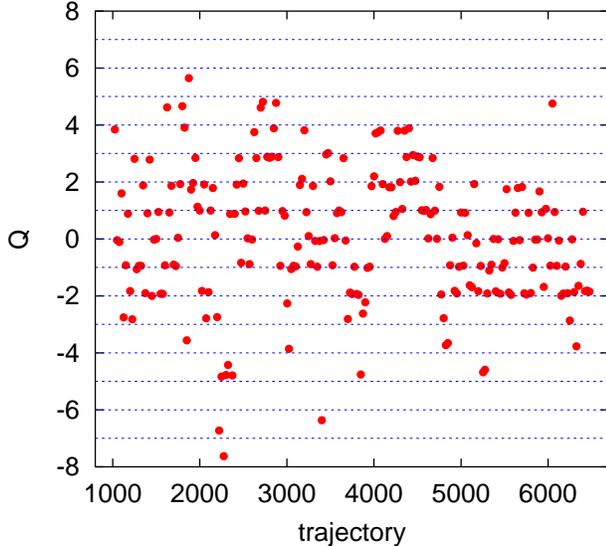}}
\caption{The topological charge $Q$ determined by the
gluonic  definition, \eq{eq:gluonic_def}, after cooling on the $16^3\times 32$
lattices at $\kappa=0.1575$. The results are concentrated around
integer values (dashed horizontal lines).}
\label{fig:cluster}
\end{figure}

In a quantum field theory, both $Q$ and $\chi$ undergo multiplicative
renormalization. Moreover, $\chi$ requires additive renormalization.
Cooling~\cite{Berg:1981nw,Teper:1985rb} is meant to filter for
(semi)-classical features; indeed, after cooling, both renormalization
constants have been shown to be close to their trivial
values~\cite{Alles:1998nu,Alles:1997nm}.  In the context of this
article we will not investigate properties of the density distribution
$q(x)$ itself but rather stick to the net topological charge $Q$ only.
For this purpose the iterative application of cooling by simple
minimisation of the Wilson plaquette action is appropriate.  To keep
the cooling update local we visit the lattice sites in an even-odd
pattern, rather than in the sequential ordering that is usually
employed in the Monte Carlo updating of pure gauge configurations.
The inner-most loop within a cooling sweep runs across the directions
$\mu$, and we carry out 60 such sweeps.  In \fig{fig:cluster} we
illustrate the resulting numerical pattern of topological charges
which cluster nicely around integer values as anticipated.
\begin{table}
\caption{Estimates of the fermionic renormalization constant $Z_P$, \eq{eq:smitvink}, assuming
the gluonic one to be unity.}
\label{tab:renorm}
\begin{tabular}{ccccc}
$\kappa$&$L_{\sigma}$&$m_0a$&$Z_Pm_0a$&$Z_P$\\\hline
0.1560&16&0.0498&0.060&1.20\\
0.1565&16&0.0395&0.056&1.43\\
0.1570&16&0.0293&0.045&1.55\\
0.1575&16&0.0192&0.029&1.49\\
0.1575&24&0.0192&0.031&1.64\\
0.1580&24&0.0092&0.017&1.79
\end{tabular}
\end{table}

The fermionic method consists of determining the topological charge
\`a la  Smit and Vink~\cite{Smit:1987fn},
\begin{equation}
\label{eq:smitvink}
Q=\nu_--\nu_+=-Z_Pm_0\tr\gamma_5M^{-1},
\label{eq:fermionic_def}
\end{equation}
that is inspired by the continuum axial divergence relation,
$\partial_{\mu}[\bar{\psi}(x)\gamma_{\mu}\gamma_5\psi(x)]=
2m\bar{\psi}(x)\gamma_5\psi(x)-2n_fq(x)$.  The above trace is to be
taken over Dirac, colour and space-time indices only (not over the
flavours), i.e.\ $\bar{\psi}\gamma_5\psi=n_f\tr\gamma_5M^{-1}$ for
$n_f$ mass degenerate quark flavours.  $Z_P = {\cal O}(1)$ is a
renormalization constant, $m_0=(\kappa^{-1}-\kappa^{-1}_c)/(2a)$
denotes the bare quark mass and $M$ is the lattice discretised version
of $D\!\!\!\!/+m_0$,
\begin{eqnarray}
2\kappa M_{xy}=\delta_{xy}
&-&\kappa\sum_{\mu}
[\left(1-\gamma_{\mu}\right)U_{x,\mu}\delta_{x+\hat{\mu},y}
\nonumber\\
&+&\left(1+\gamma_{\mu}\right)U^{\dagger}_{x-{\hat{\mu}},\mu}
\delta_{x-\hat{\mu},y}].
\end{eqnarray}
We determine $\tr\gamma_5M^{-1}$ using $Z_2$ noisy sources
with diagonal improvement as detailed in Ref.~\cite{Viehoff:1999ze}.
On the $\kappa=0.158$ configurations 100 such estimates were performed
while at all other $\kappa$ values we averaged over 400 estimates\footnote{
Approximating $Q$ by a finite number of noise vectors
can result in an
underestimated integrated autocorrelation time.
Nonetheless, we find
the $\tau_{\mbox{\scriptsize int}}$-estimates
from the fermionic definition~\cite{inprep2} to be consistent
with the gluonic ones
of Table~\ref{tab:simul}.}.

\begin{figure}[!htb]
\centerline{\epsfxsize=8cm\epsfbox{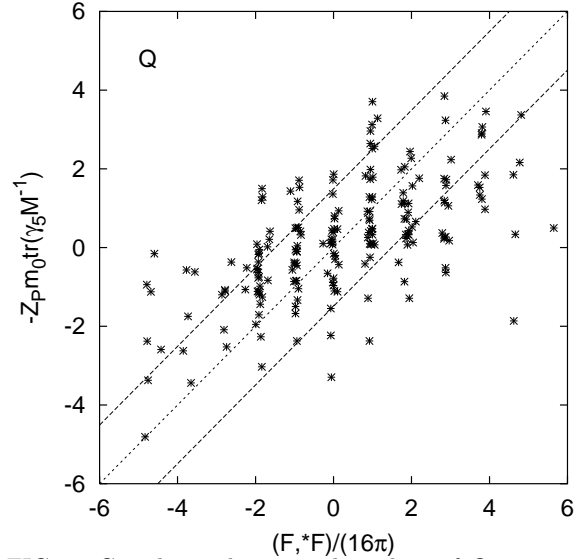}}
\caption{Correlation between the values of $Q$ as computed
from the gluonic [horizontal axis, \eq{eq:gluonic_def}]
and fermionic [vertical axis, \eq{eq:fermionic_def}, $Z_P=1.49$]
definitions on $16^3\times 32$ lattices at
$\kappa = 0.1575$. 
The width of the error band corresponds to
typical statistical uncertainties, due to the stochastic estimation of the
fermionic values. 62~\% of the data points lie within this $1\sigma$
region.}
\label{fig:compare}
\end{figure}

The renormalization constant $Z_P$ in \eq{eq:smitvink} is unknown.  We
attempt to estimate the combination $Z_Pm_0a$ from the ratio of the
gluonic [\eq{eq:gluonic_def}] and of the fermionic
[\eq{eq:fermionic_def})] definitions.  We determine $m_0a$ using
the critical value, $\kappa_c=0.15849(2)$~\cite{inprep}.  The results
are displayed in \tab{tab:renorm}, assuming the multiplicative
renormalization of the gluonic definition after cooling to be unity.
The estimates of the fermionic traces are subject to statistical
uncertainties, $\Delta Q/Z_P\approx 1$ on the $16^3\times 32$ lattices
and $\Delta Q/Z_P\approx 2$ -- 3 on the $24^3\times 40$ volumes.  In
addition, both definitions are expected to suffer from different
${\mathcal O}(a)$ lattice artefacts.  We do not attempt to estimate
the resulting statistical and systematic uncertainties on $Z_P$. We
find $Z_P$ to be of order one and to depend monotonously on the quark
mass, with the deviation from unity increasing when approaching the
chiral limit.

\begin{figure}[!htb]
\centerline{\epsfxsize=8cm\epsfbox{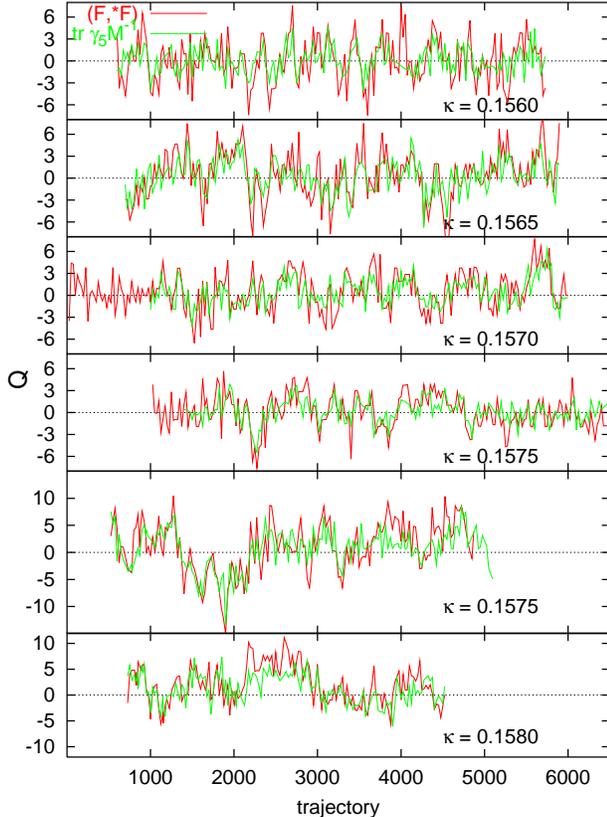}}
\caption{
 Time histories of the topological charge $Q$ for the gluonic [$(F,*F)$,
  \eq{eq:gluonic_def}] and the fermionic [$\tr\gamma_5M^{-1}$,
  \eq{eq:fermionic_def}] definitions.
  The last two series have been obtained on $24^3\times 40$ lattices, the
  first four on $16^3\times 32$ volumes.}
\label{fig:time}
\end{figure}

The correlation between the two definitions is visualised in the
scatter plot, \fig{fig:compare}, for the $16^3\times 32$ lattices at
$\kappa=0.1575$. While the gluonic data points cluster around integer
values, this is not the case for the (stochastically estimated)
fermionic values.  The data of the figure are
normalised such that the points should collapse onto the line with
slope {\it one}. Indeed, 137 out of 199 values (62~\%) lie within the
one $\sigma$ error band.  The correlation between the two definitions
is also visible from the Monte Carlo histories depicted
in~\fig{fig:time}.

\subsection{Ergodicity}
In \fig{fig:time} we survey the Monte Carlo histories of $Q$ for all
our runs.
These charts provide a gross evidence for the quality of our data in view of
the decorrelation with respect to the topological sectors.
The topological susceptibility,
\begin{equation}
\chi=\sum_x\langle q(x)q(0)\rangle=\frac{\langle Q^2\rangle}{V},
\end{equation}
should be independent of the volume $V=L_{\sigma}^3\times L_{\tau}a^4$
to a first approximation.  Therefore, the modulus of the topological
charge $\langle|Q|\rangle$ should scale in proportion to $\sqrt{V}$.
Indeed, the topological charge distribution on the large lattice at
$\kappa=0.1575$ is by a factor of about two wider than that on the
small lattice.  We also observe reduced fluctuations as we increase
$\kappa$ at fixed $\beta$.  While the $16^3\times 32$ time histories
appear to tunnel ergodically through all topological sectors the total
number of tunnellings observed for $\kappa=0.1580$ is not yet
sufficient to achieve a symmetric distribution.

\section{Topology and the hadron spectrum}
\label{sec:hadron}
We address the question whether the statistics presented in
\fig{fig:time} suffices to expose a significant $|Q|$ dependency of
hadronic states. To minimise statistical errors we subdivided each
sample into {\it two} subsamples only, one containing configurations
with $|Q|\leq 1.5$ and one with $|Q|>1.5$.  For the smaller volume at
$\kappa=0.1575$ this division results in two subsamples of
approximately 100 configurations each.

\begin{figure}[!htb]
\centerline{\epsfxsize=8cm\epsfbox{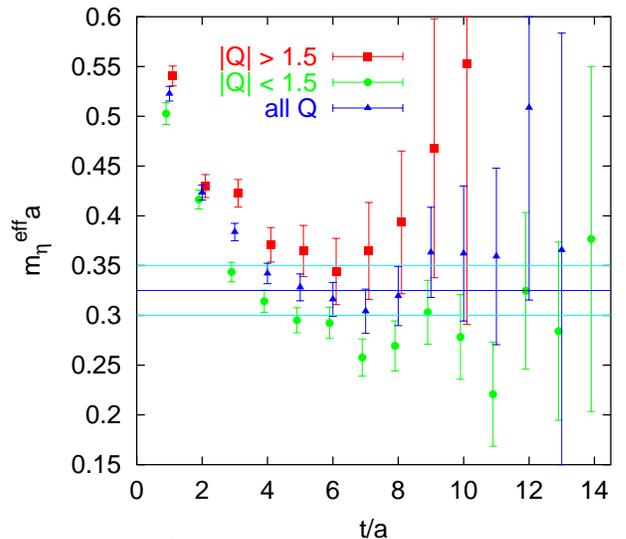}}
\caption{The $\eta'$ effective mass from a local-smeared correlation
  function on the $16^3\times 32$ lattice at $\kappa=0.1575$, separately
  determined on configurations with topological charge $|Q|\leq 1.5$ and
  $|Q|>1.5$.  The horizontal error band is the fitted asymptotic mass obtained
  on the full sample~\protect\cite{Struckmann:2000bt}.}
\label{fig:eta}
\end{figure}

The most interesting test case is given by the effective masses in the
flavour singlet pseudoscalar channel, $m_{\eta'}^{\mbox{\scriptsize
    eff}}(t)$, that should be particularly sensitive to the vacuum
topology. This is confirmed by the data displayed in \fig{fig:eta}.
We find the $\eta'$ mass on vacuum configurations with $|Q|\leq 1.5$
to systematically lie below the corresponding mass obtained with the
cut $|Q|>1.5$. Note that in the infinite volume limit one would expect
such a sensitivity of correlation functions on $|Q|$ to disappear.
For further details we refer to
Ref.~\cite{Struckmann:2000bt}.

\begin{figure}[!htb]
\centerline{\epsfxsize=8cm\epsfbox{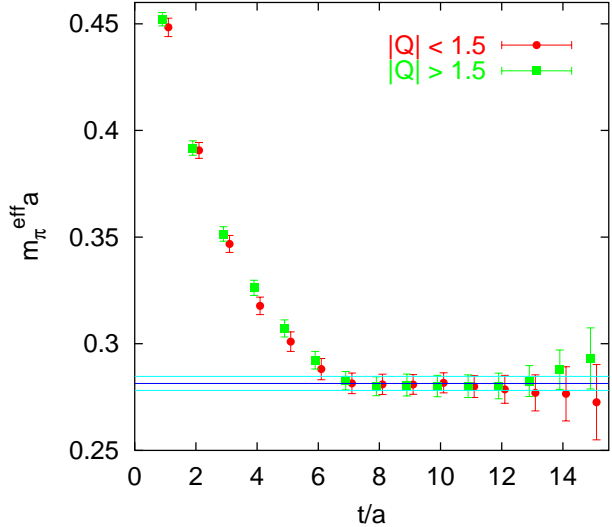}}
\caption{The same as \protect\fig{fig:eta} for the $\pi$ mass.}
\label{fig:pi}
\end{figure}

In \fig{fig:pi} we show the corresponding flavour octet channel quantities,
i.e.\ $\pi$ effective masses, where we anticipate no such correlation with topology.
This is indeed borne out by the data: not only at $t\rightarrow\infty$
but time slice by time slice no sensitivity to the value of the modulus of the
topological charge is detected, with rather high statistical accuracy.
The horizontal line with error band indicates the asymptotic large $t$
results obtained from a fit to the entire data sample~\cite{inprep}.

\begin{figure}
\centerline{\epsfxsize=8cm\epsfbox{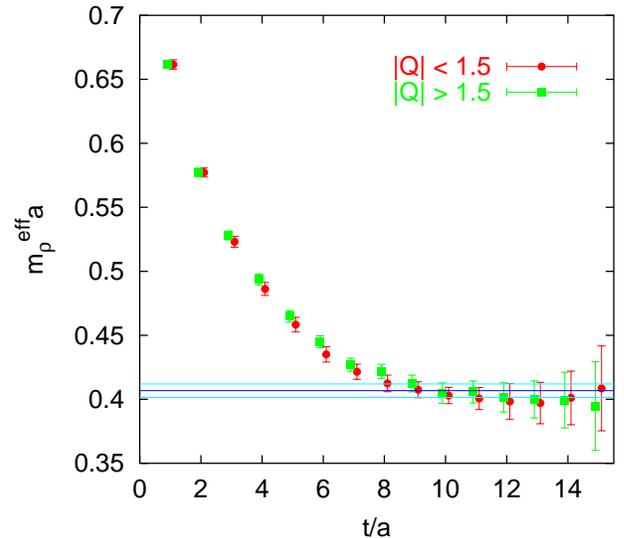}}
\caption{The same as Fig.~\ref{fig:eta} for the $\rho$ mass.}
\label{fig:rho}
\end{figure}
\begin{figure}
\centerline{\epsfxsize=8cm\epsfbox{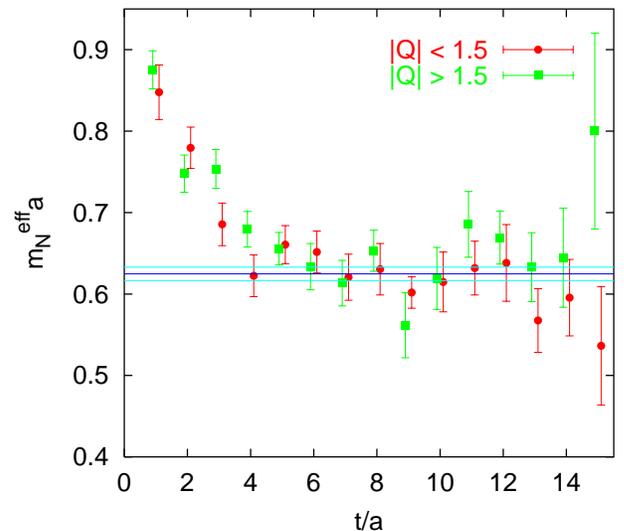}}
\caption{Nucleon effective masses from smeared-smeared correlation functions.}
\label{fig:nuc}
\end{figure}
\begin{figure}
\centerline{\epsfxsize=8cm\epsfbox{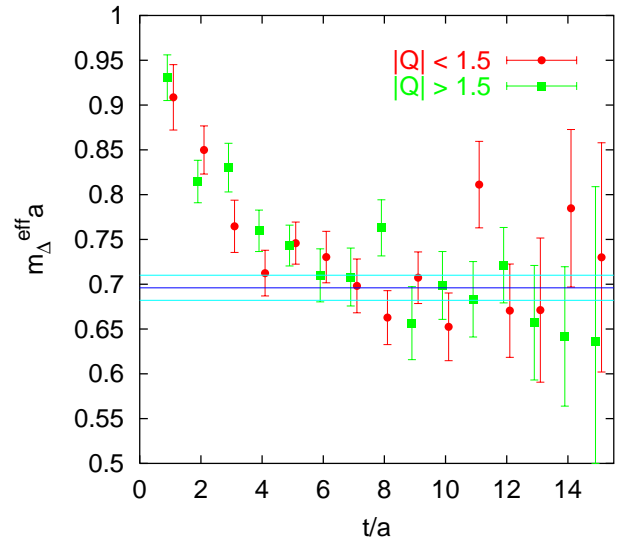}}
\caption{The same as Fig.~\ref{fig:nuc} for the $\Delta$ mass.}
\label{fig:delta}
\end{figure}

Apart from the $\eta'$ none of the standard mesonic, baryonic and
glueball-like states exhibit correlations between $|Q|$ and the
respective effective masses.  In Figs.~\ref{fig:rho} --
\ref{fig:delta} we illustrate this for the $\rho$, the
nucleon and the $\Delta$.  In no case have we found any systematic
effects on the static potential.  Of course this does not exclude the
possibility of correlations between hadronic properties and the
distribution of instantons and anti-instantons and their relative
orientations in colour and position space~\cite{Ilgenfritz:1998cd}.

\section{The topological susceptibility}
\label{sec:results}
\begin{figure}[!htb]
\centerline{\epsfxsize=8cm\epsfbox{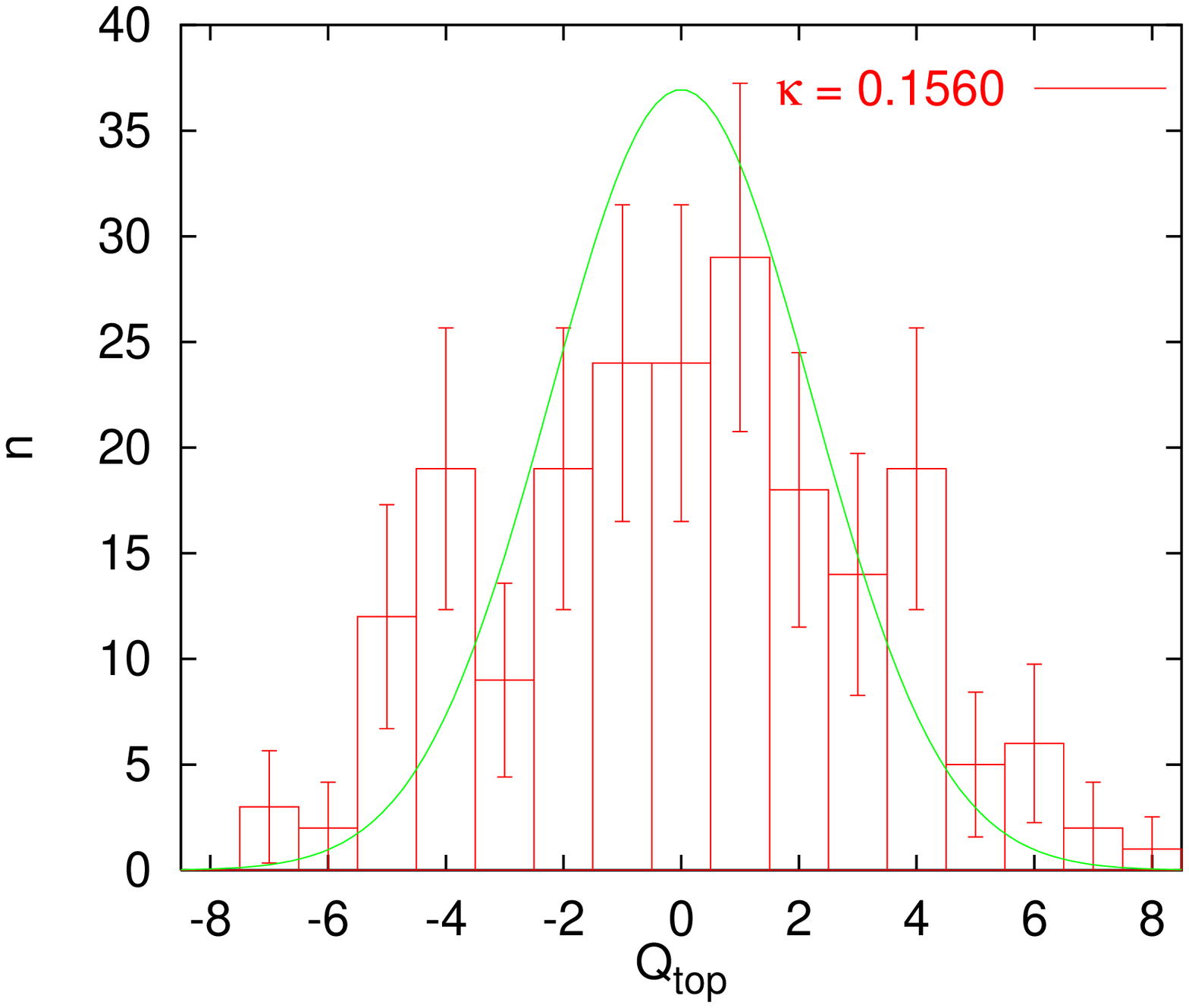}}
\caption{Topological charge distribution at $\kappa=0.156$.}
\label{fig:hist1}
\end{figure}

\begin{figure}[!htb]
\centerline{\epsfxsize=8cm\epsfbox{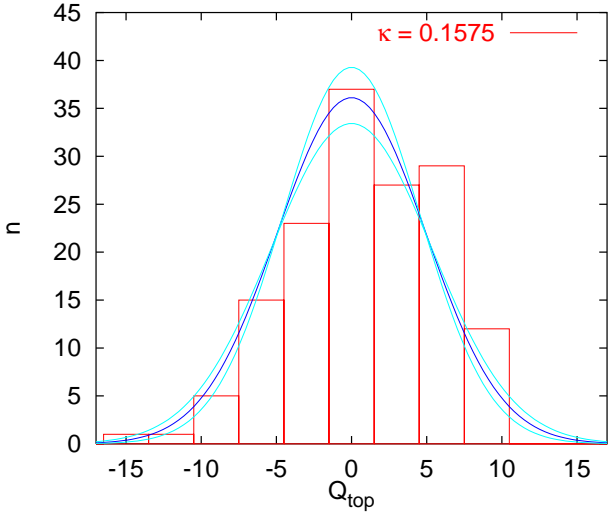}}
\caption{The same as Fig.~\protect\ref{fig:hist1} for
$\kappa=0.1575$, $L_{\sigma}=24$.}
\label{fig:hist2}
\end{figure}

\begin{figure}[!htb]
\centerline{\epsfxsize=8cm\epsfbox{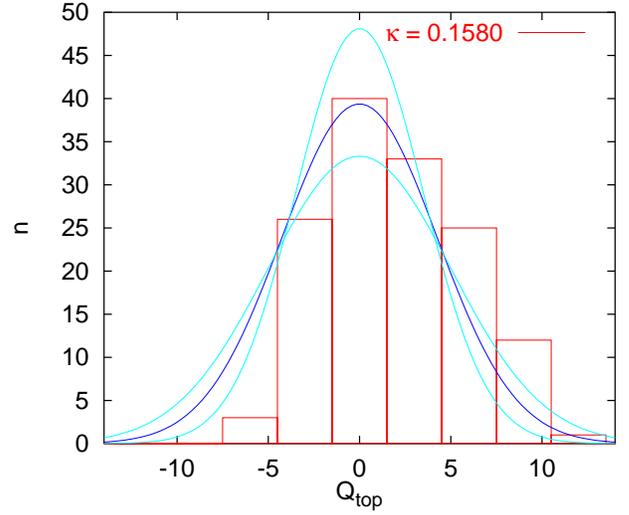}}
\caption{Topological charge distribution at $\kappa=0.158$.}
\label{fig:hist3}
\end{figure}

In Figs.~\ref{fig:hist1} -- \ref{fig:hist3} we display three
histograms of topological charge distributions.  Each bin with width
$\delta Q$, centred around $Q_j=j\delta Q$, contains all measurements
resulting in charges within the interval $[Q_j-\delta Q/2,Q_j+\delta
Q/2]$, where $\delta Q=1$ on the $16^3\times 32$ lattice depicted in
\fig{fig:hist1} and $\delta Q=3$ for the $24^3\times 40$ lattices of
\fig{fig:hist2} and \fig{fig:hist3}.  In addition to the data we
display Gaussian distributions,
\begin{equation}
n(Q)=\frac{n_{\mbox{\scriptsize meas}}\delta Q}{
\sqrt{2\pi {\langle Q^2\rangle} }}  \exp
\left(-\frac{Q^2}{2\langle Q^2\rangle}\right).
\end{equation}
In \fig{fig:hist1} we include the statistical uncertainties of the
individual bins, $\Delta n= 2\tau_{\mbox{\scriptsize
    int}}\nu_{\mbox{\scriptsize meas}}\sqrt{n}$, while in
\fig{fig:hist2} and \fig{fig:hist3} the error on the width of the
distribution is reflected by the error band around the central curve.

\begin{table}
\caption{Topological susceptibilities.}
\label{tab:topsus}
\begin{tabular}{cccccc}
$\kappa$&$L_{\sigma}$&$\langle Q\rangle$&$\chi a^4/10^{-6}$&$\chi^{1/4}r_0$&
$\chi^{1/4}/$MeV\\\hline
0.1560&16&0.04(32)&69(6)&0.466(11)&183(4)(9)\\
0.1565&16&0.48(48)&83(13)&0.505(22)&199(8)(10)\\
0.1570&16&0.60(34)&57(10)&0.476(22)&187(8)(9)\\
0.1575&16&0.05(33)&41(7)&0.472(21)&186(8)(9)\\
0.1575&24&0.84(1.33)&44(7)&0.479(20)&189(8)(10)\\
0.1580&24&1.92(1.01)&31(12)&0.465(44)&183(17)(9)\\\hline
quenched&16&-0.68(33)&80(14)&0.505(22)&199(9)(10)
\end{tabular}
\end{table}

The fitted topological susceptibilities $\chi=\langle Q^2\rangle/V$ as well as
$\langle Q\rangle$ are displayed in \tab{tab:topsus}.  For orientation, we
convert $\chi$ into physical units in the last column of the table, using
$r_0^{-1}=(394\pm 20)$~MeV.  The distribution at $\kappa=0.158$ is not
symmetric around zero anymore, as reflected by the value $\langle
Q\rangle=1.9\pm 1.0$ and by \fig{fig:hist3}. Therefore, in this case,
the resulting value of $\chi$ should pass  with a grain of doubt,
notwithstanding the  comfortable error bars.  A comparison between the
$L_{\sigma}=16$ and $L_{\sigma}=24$ results at $\kappa=0.1575$ reveals that
the level of finite size effects on $\chi$ is  below the statistical errors.

The Gell-Mann-Oakes-Renner (GMOR) relation connects the pion mass to
the chiral quark mass, $m$, via the pion decay constant\footnote{Our
  normalisation conventions are such that $f_{\pi}\approx 92$~MeV.}
$f_{\pi}$ in the chiral limit and the chiral condensate, $-\Sigma$:
\begin{equation}
\label{eq:gmor}
f_{\pi}^2m_{\pi}^2=2m\Sigma+{\mathcal O}(m^2).
\end{equation}
A direct lattice determination of $f_{\pi}$ from the appropriate
current matrix element yields $\sqrt{2}f_{\pi}= Z_A\langle
0|A_4|\pi\rangle/m_{\pi}\rightarrow 0.0496(34)a^{-1}$, in the limit
$m_{\pi}^2\rightarrow 0$~\cite{Eicker:1999sy}.  Allowing for an
additional systematic error of 10~\% on $f_{\pi}$ to account for
the fact that we have only determined the
axial vector renormalization constants $Z_A$
perturbatively, we arrive at the lattice
estimate
\begin{equation}
\label{eq:fpi}
f_{\pi}=(0.238\pm 0.030)\,r_0^{-1}=(94\pm 13)\,\mbox{MeV}
\end{equation}
at $\kappa=\kappa_c$, using the value
$r_0^{-1}a=0.1476^{+47}_{-30}$~\cite{Bali:2000vr}.  In combining
\eq{eq:gmor} with \eq{eq:chiral} one expects\footnote{This formula as
  well as the GMOR relation \eq{eq:gmor} only apply to $n_f\geq 2$
  since for $n_f=1$ no light pion exists (for a finite number of
  colours, $N_c<\infty$).}
\begin{equation}
\label{eq:extra}
\chi=\frac{f_{\pi}^2m_{\pi}^2}{2n_f}+{\mathcal O}(m_{\pi}^4)
\end{equation}
for small pseudoscalar masses
and large Leutwyler-Smilga parameters, 
$x=Vm\Sigma\gg 1$~\cite{Leutwyler:1992yt,Durr:2001ei}.

\begin{table}
\caption{The spatial lattice extents in units of the pion correlation length,
the Leutwyler-Smilga parameters, $x=Vm\Sigma$ and topological charge
fluctuations $\langle Q^2\rangle$.}
\label{tab:leut}
\begin{tabular}{ccccc}
$\kappa$&
$V/a^4$&
$m_{\pi}aL_{\sigma}$&
$x$&
$\langle Q^2\rangle$\\
\hline
0.1560&$16^3\times 32$&7.14(4)&$16.0\pm 4.1$&$9.1\pm 0.8$\\
0.1565&$16^3\times 32$&6.39(6)&$12.9\pm 3.2$&$10.9\pm 1.7$\\
0.1570&$16^3\times 32$&5.51(4)&$9.6\pm  2.4$&$7.5\pm 1.3$\\
0.1575&$16^3\times 32$&4.50(5)&$6.4\pm  1.6$&$5.4\pm 0.9$\\
0.1575&$24^3\times 40$&6.65(6)&$26.1\pm 6.6$&$24.1\pm 3.9$\\
0.1580&$24^3\times 40$&4.77(7)&$13.4\pm 3.4$&$17.2\pm 6.6$
\end{tabular}
\end{table}

\begin{figure}[!htb]
\centerline{\epsfxsize=8cm\epsfbox{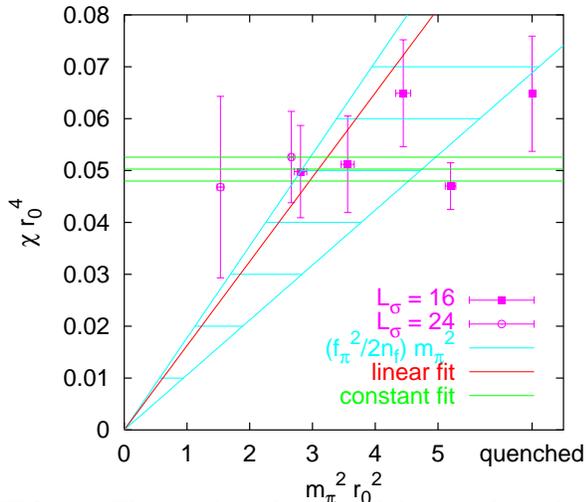}}
\caption{The topological susceptibility in physical units as a function of $m_{\pi}^2$.
  The right-most data point is from the quenched reference simulation
  and has not been included into the fits. The dashed error band
  corresponds to the small mass expectation from the independently
  determined pion decay constant $f_{\pi}=0.238(30)\,r_0^{-1}$.}
\label{fig:chi}
\end{figure}

In Table~\ref{tab:leut} we display the
spatial lattice extents, $L_{\sigma}am_{\pi}$ as well as $x$
and $\langle
Q^2\rangle=\chi V$. We estimate
the (scheme independent) combination $m\Sigma$ that appears within $x$
{}from
Eqs.~(\ref{eq:gmor}) -- (\ref{eq:fpi}). The topological
charge fluctuations $\langle Q^2\rangle$ correspond to the average number of
non-interacting instantons that can be accomodated within the
simulated lattice volume~\cite{Schafer:1998wv,Durr:2001ty}. 
It has
been argued~\cite{Durr:2001ty} that $x>10$ is already a large volume in the
Leutwyler-Smilga~\cite{Leutwyler:1992yt}
sense, such that Eq.~(\ref{eq:extra}) should safely
apply to all but the $16^3\times 32$ lattices at $\kappa=0.157$ and
$\kappa=0.1575$. Even at the latter, more critical, $\kappa$-value,
where we have two lattice volumes that correspond to
$x\approx 6$ and $x\approx 25$ at our disposal, we do not
find any volume dependence of $\chi$.

In the regime of large $m_{\pi}$ one would expect $\chi$ in
units of some reference scale like $r_0^{-1}$ to approach the quenched
value as a smooth function of $m_{\pi}$.  Prior to the comparison with
\eq{eq:extra} one should in principle extrapolate the lattice results
on $\chi$ to the continuum limit, for instance along a trajectory of
constant $m_{\pi}r_0$. Having only $\beta=5.6$ data at our disposal we
cannot yet perform this extrapolation but conjecture that our results
are already sufficiently close to the continuum limit for
\eq{eq:extra} to hold.  This assumption is plausible in view of the
fact that the topological charges obtained from the gluonic and
fermionic definitions agree with each other reasonably well (with a
renormalization constant $Z_P$ of order one) and of our previous
studies of light hadronic quantities~\cite{Bali:2000vr,Eicker:1999sy}.

For the purpose of the present investigation we shall consider the
dimensionless topological susceptibility, $\chi r_0^4$, as a function
of the dimensionless combination $m_{\pi}^2r_0^2$ in order to carry
out the chiral extrapolation.  The results are plotted in
\fig{fig:chi}, together with the region that is allowed for by the
leading order expectation, \eq{eq:extra}, for the $f_{\pi}$ value of
\eq{eq:fpi}.  All unquenched data except for the point at the heaviest
quark mass ($m_{\pi}^2r_0^2\approx 5$) are consistent with this expectation.
Note that the most chiral data point stems from the run with
$\kappa=0.158$ which, as can be seen from Figs.~\ref{fig:time} and
\ref{fig:hist3}, is not of sufficient statistical quality for the
purpose of topological studies.  The right-most entry in \fig{fig:chi}
is the result of our quenched reference study.

A linear fit, excluding the largest mass
point, to the parametrisation~(\ref{eq:extra})
renders $f_{\pi}=0.255(11)r_0^{-1}$ with $\chi^2/N_{DF}=3.93/4$.
This value compares reasonably well with the expectation,
\eq{eq:fpi}. On the other hand our data are consistent
with a mass independence of $\chi$ too: fitting them to a constant yields
$\chi\, r_0^4=0.0503(23)$ with $\chi^2/N_{DF}=2.67/5$, in agreement with the
quenched reference point, $\chi\, r_0^4=0.065(11)$.

Our present study is based on data obtained at a fixed value of
$\beta$ and the quark mass is varied by just tuning $\kappa$. It is
worthwhile to relate our findings to those recently presented by the
UKQCD collaboration\footnote{Unfortunately,
none of the other collaborations
have converted their results into units of $r_0$.}~\cite{Hart:2000hy}
 who simultaneously vary $\beta$
and $\kappa$ to change the quark mass while keeping
the lattice spacing $a$ fixed in
units of $r_0$. Apart from this they dispose of ensembles
of gauge field configuration of comparable statistical sizes, using
slightly coarser lattice spacings with an improved fermionic action.

\begin{figure}[!htb]
\centerline{\epsfxsize=8cm\epsfbox{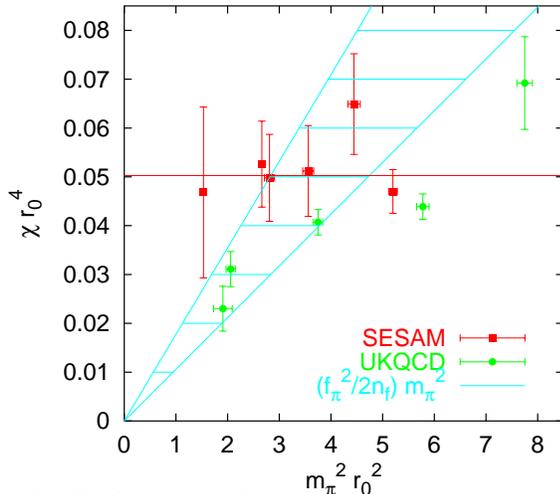}}
\caption{Our results in comparison to those reported by the UKQCD
Collaboration~\protect\cite{Hart:2000hy}.
The horizontal line is a constant fit to our data.
The dashed error band
corresponds to the small mass expectation from the
pion decay constant.}
\label{fig:ukqcd}
\end{figure}

In order to facilitate this comparison we have superimposed both data
sets\footnote{The statistical errors quoted by the UKQCD
  collaboration are somewhat smaller than ours since no signs of
  autocorrelation effects have been found in their
  study~\cite{hart}.}
in \fig{fig:ukqcd}. The data do not clearly
contradict each other, however, we are unable to confirm the decrease of
the topological susceptibility that has been reported in
Ref.~\cite{Hart:2000hy},
at pion masses within the range of
present-day lattice simulations.

\section{Conclusions}
We have demonstrated that Wilson fermions are suitable for lattice
studies of topology related effects in the QCD vacuum. We have found
agreement between the topological charge computed after cooling from
the field strength tensor, $Q=(F,*F)/(16\pi^2)$, and the fermionic
definition given by Smit and Vink with a renormalization constant
$Z_P$ of order one.  The trace has been estimated from diagonal
improved $Z_2$ noisy sources.  As the next step we plan to investigate
correlations between the topological charge density distribution and
eigenvectors of $\gamma_5M$.

The topological susceptibility at $\beta=5.6$ has been found to be
consistent with that of quenched studies at large sea quark masses.
Our data at smaller mass values is consistent
with the asymptotic slope in $m_{\pi}^2$ that is expected from 
our independently determined $n_f=2$ pion
decay constant. However, we are unable to unambiguously confirm
the decrease that has
recently been reported by the UKQCD collaboration~\cite{Hart:2000hy}
with a different fermionic action.
Results obtained by the CP-PACS collaboration~\cite{AliKhan:2000zi}
who cover a similar range of quark masses on somewhat coarser lattices
do not clearly show this tendency either.
We plan to clarify this issue in
simulations at different $\beta$ values which will enable us to perform
a continuum limit extrapolation, and at smaller sea quark masses.

\acknowledgements We thank S.\ G\"usken and P.\ \"Uberholz for their
contributions at an earlier stage of this project.
We gratefully acknowledge R.\ Burkhalter for detecting an important
mistake in this article. Our European
collaboration was funded by the EU network ``Hadron Phenomenology from
Lattice QCD'' (HPRN-CT-2000-00145).  G.B.\ received support from EU
grant HPMF-CT-1999-00353 and PPARC grant
PPA/G/O/1998/00559, N.E.\ from DFG grant Li~701/3-1.
B.O.\ appreciates support from the DFG
Graduiertenkolleg ``Feldtheoretische und Numerische Methoden in der
Statistischen und Elementarteilchenphysik''.  The HMC productions were
run on the APE100 systems at INFN Roma and NIC Zeuthen.  We are
grateful to our colleagues G.\ Martinelli and F.\ Rapuano for the
fruitful T$\chi$L-collaboration. Part of the physics evaluation has
been carried out on the ALiCE cluster computer at Wuppertal University.

\end{document}